\documentclass[prl,aps,twocolumn, floatfix, superscriptaddress, showpacs]{revtex4}
\usepackage{amsmath,bm,graphicx}
\usepackage{amssymb}
\usepackage{amsfonts}

\newcommand{\V}[1]{{\bf #1}}
\newcommand{\so}{\sigma_1}
\newcommand{\st}{\sigma_2}

\begin{document}

\title{On the probability distribution of power fluctuations in turbulence.}

\author{M. M. Bandi}
\email[Corresponding Author: ]{mbandi@lanl.gov}
\affiliation{Center for Nonlinear Studies / MPA10, LANL, Los Alamos, NM 87545, USA}
\author{Sergei G. Chumakov}
\affiliation{Center for Nonlinear Studies / T3, LANL, Los Alamos, NM 87545, USA}
%\email{chumakov@lanl.gov}
\author{Colm Connaughton}
%\email{connaughtonc@gmail.com}
\affiliation{Mathematics Institute and Centre for Complexity Science, University of Warwick, Coventry CV4 7AL, UK}

\date{\today}

\begin{abstract}
We study local power fluctuations in numerical simulations of stationary, homogeneous, isotropic turbulence in two and three dimensions with Gaussian forcing. Due to the near-Gaussianity of the one-point velocity distribution, the probability distribution function (pdf) of the local power is well modelled by the pdf of the product of two joint normally distributed variables. In appropriate units, this distribution is parameterised only by the mean dissipation rate, $\epsilon$. The large deviation function for this distribution is calculated exactly and shown to satisfy a Fluctuation Relation (FR) with a coefficient which depends on $\epsilon$. This FR is entirely statistical in origin. The deviations from the model pdf are most pronounced for positive fluctuations of the power and can be traced to a slightly faster than Gaussian decay of the tails of the one-point velocity pdf. The resulting deviations from the FR are consistent with several recent experimental studies.
\end{abstract}

%\keywords{}
\maketitle

We study the pdf of local power fluctuations in two--dimensional (2D) and three--dimensional (3D) turbulence, important practical examples of strongly non--equilibrium stationary states. Stationary turbulence requires external forcing to counter viscous dissipation producing a balance of the average rates of energy injection (power) and dissipation. The power is locally a scalar product of the force and velocity. The latter always has intrinsic stochasticity. The power thus has non-trivial statistics of its own. Interest in the statistics of the power comes  from two principal directions. From an engineering perspective, the average power relates directly to the drag on a body in a turbulent flow. From a theoretical perspective, interest focuses primarily on the power fluctuations. Such non-equilibrium fluctuations get to the heart of the differences between equilibrium and non-equilibrium statistical mechanics as they relate directly to the lack of detailed balance in turbulence.

Experimental studies of the input power in turbulence initially focused on the mean and its scaling with Reynolds  number \cite{LFS1992}. The subsequent realisation that certain types of non-equilibrium fluctuations exhibit an exact symmetry known as a Fluctuation Relation (FR) (see \cite{CCJ2006} and the references therein) has focused attention on fluctuations about the mean in non--equilibrium systems. \cite{LPF1996}. Turbulence has been harnessed as a source of such fluctuations in various contexts \cite{AFMP2001}. Specific studies of the power pdf have recently been undertaken for wave turbulence \cite{FLF2008} and 2D turbulence \cite{BC2008}. It was shown that the pdf of power fluctuations in different turbulent systems can be qualitatively modelled by the pdf of the product of two joint normally distributed variables, $v$ and $f$, the velocity and force respectively. 

In this Letter, we consider the statistics of the power in 2D and 3D turbulence with Gaussian external forcing. We show that a product of normal variables captures the qualitative features of the pdf in both cases. We calculate the large deviation function (Kramer function) and find an exact FR with a rate depending on $\rho$, the correlation coefficient of the two variables. For turbulence, $\rho$  is proportional to the mean dissipation rate. This is entirely a consequence of statistics and has no relation to the dynamical arguments underlying some theoretical results.  This may partially explain the ubiquity of experimental FRs in the literature and the lack of agreement on the value or meaning of the measured rate (see \cite{BCG2008} and the references therein for discussion of FR experiments). Applied to turbulence, this model, while qualitatively appealing, does not correctly capture the far positive tail of the power pdf. This is traced to slightly faster than Gaussian decay of the one-point velocity distribution. This is in accordance with theoretical expectations and results in a deviation of the FR from the linear scaling which is consistent with the results of several experiments \cite{BCG2008,FLF2008}.

We solve the 2D and 3D incompressible Navier--Stokes equations for the velocity, $\V{v}(\V{x},t)$, with a time--independent force, $\V{f}(\V{x})$ and bulk drag term, $\alpha \V{v}$:
\begin{eqnarray}
\label{eq-NS}  \partial_t {\V{v}} + (\V{v} \cdot \nabla)\,\V{v} &=& -\nabla p + \nu \Delta \V{v} -\alpha \V{v} + \V{f}\\
\nonumber \nabla\cdot \V{v} &=& 0
\end{eqnarray}
Stationarity requires finite $\alpha$ for 2D flows where dissipation of energy transferred to large scales by the inverse cascade is needed. $\alpha=0$ for 3D flows since there is no inverse cascade. Our simulations were done in biperiodic domains using standard pseudo-spectral methods. For numerical details, see \cite{BC2008} (2D) and \cite{CHU2006} (3D). The forcing is central in what follows so let us clarify the detail. Unlike the temporally-decorrelated  forcing often used to drive simulations of isotropic turbulence, our forcing has no time-dependence.  It does have spatial disorder. It is generated by selecting modes in a shell, $k_1 < \left|\V{k}\right| < k_2$, in the space of wave-vectors, $\V{k}$. These are assigned an ampitude, $A(\left|\V{k}\right|)$ and a random phase uniformly distributed on $\left[0, 2\pi \right)$.  We took $A(\left|\V{k}\right|)$ to be the indicator function on $\left[ k_1, k_2\right]$. We project out the non-solenoidal component to assure incompressibilty. An inverse Fourier transform then produces a spatially random forcing field. In the 2D simulations, $\V{f}$ has a single component: $\V{f}_{\rm 2D}(\V{x}) = (0,f_2(\V{x}))$. The current was applied in the $\V{x}$ direction and the magnetic field is perpendicular to the fluid layer so that the Lorentz force acts purely in the $\V{y}$ direction (see \cite{BC2008}). This simplifies things but is not an essential point. Indeed, in the 3D simulations, all three components of the force were present:  $\V{f}_{\rm 3D}(\V{x}) = (f_1(\V{x}),f_2(\V{x}), f_3(\V{x}))$. 

\begin{figure}[t]
\includegraphics[width=6.5cm]{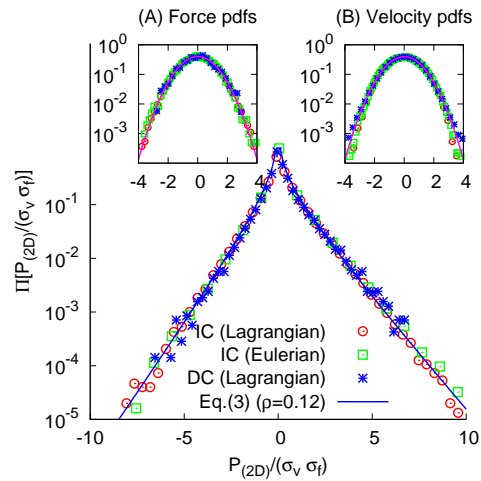}
\caption{(Color online) \label{fig-power2D}Pdfs of the 2D power normalized by $\sigma_v \sigma_f$ for the inverse ($\circ$) and direct ($\ast$) cascades in the Lagrangian frame and the inverse cascade in the Eulerian frame (${\scriptsize \Box}$) . The solid line is Eq. (\ref{eq-XYDistribution}). The insets show the corresponding pdfs of $\V{f}$ (A)  and $\V{v}$ (B) normalized by their standard deviations. }
\end{figure}

\begin{figure}
\includegraphics[width=6.5cm]{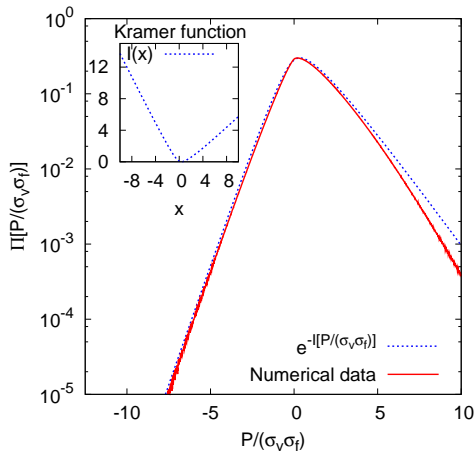}
\caption{\label{fig-totalPower3D}(Color online) Comparison of the pdf of the  local 3D power with $e^{-I(x)}$. Inset shows the Kramer function, $I(x)$, given by Eq~(\ref{eq-kramerFunction}), for $\rho=0.35$.}
\end{figure}

The rationale for this forcing is two-fold. Firstly, our 2D forcing exactly mimics that  used to generate turbulence in electromagnetically driven fluid layers \cite{PT1997,RDE2005}.  It is thus of direct relevance to 2D experiments.  Secondly, since we are interested in power fluctuations, it is attractive to limit the sources of stochasticity to  the intrinsic randomness of the turbulent fluctuations. By the Central Limit Theorem, our forcing protocol produces a Gaussian distribution for the single-point pdf of $\V{f}$ provided that enough modes participate. This is shown in the inset of Fig.~\ref{fig-power2D} for the 2D case and of Fig.~\ref{fig-totalPower3D} for the 3D case. We should be clear that we are not attempting to make any universal statements. Although Gaussian forcing is often used in numerical simulations and has experimental relevance, it has no a-priori justification.

Turning to the velocity, $\V{v}$, its single point pdf is known to be close to Gaussian for homogeneous, isotropic turbulence since the early days of turbulence theory \cite{MY1975}. On the other hand, the Navier--Stokes equation, Eq.~(\ref{eq-NS}) is nonlinear. Even with Gaussian forcing, there is no reason to expect that the pdf of $\V{v}$ should be exactly Gaussian and indeed it is not. While most investigations have focused on the relatively large non--Gaussianity of velocity differences, careful measurements show that the single-point pdf of $\V{v}$ decays slightly faster than Gaussian in both the 2D \cite{JWZ2006} and 3D \cite{GFN2002} cases. 

We now consider the local power, denoted by $p$. The 2D power is a simple product, $p_{\rm (2D)} = v_y f_y$. The 3D power has three contributions: $p_{\rm (3D)} = \sum_{i=1}^3 v_i f_i$. Ignoring for now any sub-Gaussian tails of the pdf of $\V{v}$, it is clear that modeling $p$ using products of Gaussian force and velocity components should capture the qualitative features of the single point pdf. This has already been proposed in a Lagrangian setting in 2D turbulence \cite{BC2008} and in the context of wave turbulence \cite{FLF2008} and shown to work very well.  In the present Letter, we extend the description to 3D flows, calculate the large deviation properties of the model and address the meaning of the FR for turbulent power fluctuations

We need some results on products of normal variables.
If $x_1$ and $x_2$ are two joint normally distributed random variables with mean zero, variances $\so^2$ and $\st^2$ and correlation coefficient $\rho$, then
their joint pdf is
\begin{equation}
\label{eq-jointDistribution}
{\mathbb P}(x_1,x_2) = \frac{1}{2 \pi \so\st\sqrt{1-\rho^2}}\ {\rm e}^{-\frac{1}{2(1-\rho^2)}\left( \frac{x_1^2}{\so^2} - 2 \rho\frac{x_1 x_2}{\so\st} + \frac{x_2^2}{\st^2} \right)}.
\end{equation}
$x_1$ and $x_2$ should be thought of as components of $\V{v}$ and $\V{f}$ respectively.  The pdf of the product, $z=x_1 x_2$, is
\begin{equation}
\label{eq-XYDistribution}
{\mathbb P}(z) = \frac{{\rm e}^{\frac{\rho z}{(1-\rho^2)\so\st}}}{\pi \so\st\sqrt{1-\rho^2}}\,{\rm K}_0\left(\frac{\left| z \right|}{(1-\rho^2)\so\st}\right)
\end{equation}
where ${\rm K}_0(z)$ is the modified Bessel function of the second kind of order zero. 
We take $\langle . \rangle$ to denote averaging with respect to the pdf, Eq.~(\ref{eq-XYDistribution}). The moment generating function, $\chi(\theta) = \langle{\rm e}^{\theta z}\rangle$ can be calculated explicitly:
\begin{equation}
\label{eq-MGF}
\chi(\theta) = \frac{1}{\sqrt{1 - 2\rho \so\st \theta - (1-\rho^2)\so^2\st^2\theta^2}}
\end{equation}
where $\theta \in (-\frac{1}{\so\st(1-\rho)},\frac{1}{\so\st(1+\rho)})$.  It is then easy to obtain moments. The mean, variance and skewness are:
\begin{eqnarray}
\label{eq-meanXY}\langle z \rangle &=& \rho\, \so\st\\
\label{eq-varianceXY}\langle z^2\rangle - \langle z \rangle^2 &=& (1+\rho^2)\,\so^2\st^2\\
\label{eq-skewnessXY}\frac{\langle z^3 \rangle - 3 \langle z \rangle \langle z^2\rangle + 2 \langle z \rangle^3}{(\langle z^2\rangle - \langle z \rangle^2)^{3/2}} &=& \frac{2 \rho \,(3+\rho^2)}{(1+\rho^2)^{3/2}}.
\end{eqnarray}
Normalising $\V{v}$ and $\V{f}$ by their standard deviations, $\sigma_v$ and $\sigma_f$, the mean of the pdf Eq.~(\ref{eq-XYDistribution}) gives the correlation coeffient, $\rho$.  For stationary turbulence, this relates $\rho$ to the average dissipation rate, $\varepsilon$. Only a single component of $\V{f}$ contributes to the 2D power so $\varepsilon=\rho$. All components contribute to the 3D power so $\varepsilon= 3\rho$. Eqs. (\ref{eq-meanXY}) -- (\ref{eq-skewnessXY}) thus link the statistics of $p$ to the dissipation rate.

\begin{figure}[t]
\includegraphics[width=6.5cm]{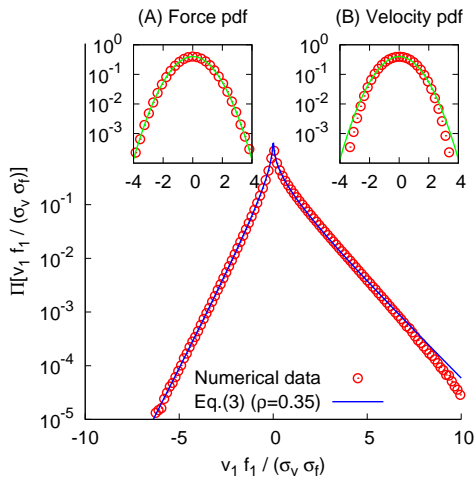}
\caption{\label{fig-power3D}(Color online) Comparison of the pdf of the $\V{v}_1\V{f}_1$ contribution to the 3D power with Eq.~(\ref{eq-XYDistribution}). Insets show the pdfs of $\V{f}_1$ (A)  and $\V{v}_1$ (B) normalized by $\sigma_v$ and $\sigma_f$.}
\end{figure}

The value of $\rho$ can be measured. A comparison between Eq.~(\ref{eq-XYDistribution}) for the measured value of $\rho$  and the contribution of a single component of the force to the power (normalised by the product of $\sigma_v$ and $\sigma_f$) is shown for   the 2D power in Fig.~\ref{fig-power2D} and for the 3D power in Fig.~\ref{fig-power3D}. 2D results are presented for both direct and inverse cascade regimes with nominal (integral scale) Reynolds numbers of 1100 and 7000 respectively. The $\rho$ values are $0.11$ and $0.13$. The (Taylor microscale) Reynolds number of the 3D simulation was 35 and $\rho$ was $0.35$. As expected, the agreement is good. In detail, the 3D simulations show a systematic deviation for large positive fluctuations. We will return to this later.

We now calculate the large deviation properties of the model pdf, Eq.~(\ref{eq-XYDistribution}). Let us briefly explain what this means and why it is useful. Suppose we take $n$ {\em independent} samples from the distribution Eq.~(\ref{eq-XYDistribution}), denoting them by $z_i$, $i=1\ldots n$. The large deviation principle for Eq.~(\ref{eq-XYDistribution}) concerns the pdf of their average, $M_n = \frac{1}{n} \sum_{i=1}^n z_i$. It states that there exists a function, $I(x)$, the rate function or Kramer function, such that 
\begin{equation}
\label{eq-LDP}
{\mathbb P}(M_n > x) \asymp {\rm e}^{- n I(x)}.
\end{equation}
This is useful for several reasons. Firstly, the 3D power is a sum of 3 random variables with distribution Eq.~(\ref{eq-XYDistribution}) so Eq.~(\ref{eq-LDP}) provides partial information about the tails of the distribution of the total power in 3D. Secondly experiments often measure global - or at least coarse--grained - power rather than local power. Eq.~(\ref{eq-LDP}) provides a link between the local and global power which may be more accessible experimentally. Finally, a FR expresses a particular symmetry of the rate function for a stochastic process, so knowing $I(x)$ allows us to address the question of a FR for Eq. (\ref{eq-XYDistribution}) directly. In this case, it is possible to obtain $I(x)$ in closed form from the Chernoff formula \cite{LR1997}: $I(x) = \max_{\theta} \left\{ \theta x  - \ln \chi(\theta)\right\}$. Lengthy but straightforward calculations yield:
\begin{widetext}
\begin{equation}
\label{eq-kramerFunction}
I(x) = \frac{(\rho^2\!-\!1)\so\st - 2 \rho x +\sqrt{4x^2 + (1\!-\!\rho^2)^2\so^2\st^2} - 2 (1\!-\!\rho^2)\so\st \ln\left[\frac{2 x^2}{\so\st((\rho^2\!-\!1)\so\st + \sqrt{4x^2 + (1\!-\!\rho^2)^2\so^2\st^2})}\right]}{2 \so\st (1\!-\!\rho^2)}.
\end{equation}
\end{widetext}

This unwieldy expression is plotted in the inset of Fig.~\ref{fig-totalPower3D} for $\rho=0.35$ and $\so=\st=1$. The main part of Fig.~\ref{fig-totalPower3D} illustrates how the asymptotic expression Eq.~(\ref{eq-LDP}) captures the essential features of the pdf of the 3D power. One cannot expect exact correspondence for several reasons. Firstly, we have seen that Eq.~(\ref{eq-XYDistribution}) over-estimates the probability of large positive values of each individual contribution to the total 3D power, an effect which remains evident when these contributions are summed. Secondly the components $\V{v}$ are not strictly independent owing to the incompressibility condition. Finally, one should remember that Eq.~(\ref{eq-LDP}) is an asymptotic statement. These objections notwithstanding, the correspondence is good.

We now turn to the question of a FR for turbulent power fluctuations. A FR
is a symmetry of the pdf of a quantity, $X_\tau$,
derived from the entropy production or energy dissipation in a non--equilibrium system. $X_\tau$ is obtained by averaging a physical
quantity, $x(t)$, typically the entropy
produced or energy dissipated over
a time interval $[t,t+\tau]$: $X_\tau = \tau^{-1}\ \int_t^{t+\tau} x(t^\prime)\ dt^\prime$.
$X_\tau$ is positive on average but, may fluctuate
sufficiently that negative fluctuations
are observable. A FR quantifies the relative
probability of a negative fluctuation over a time interval
compared to the probability of a positive fluctuation of the same
magnitude. The ratio of probabilities takes the form:
\begin{equation}
\label{eq-generalFR}
\frac{\Pi(X_\tau)}{\Pi(-X_\tau)} = e^{\Sigma\,\tau\,X_\tau},
\end{equation}
where $\Sigma$ is a constant, independent of the averaging interval, $\tau$. Clearly this equates to the rate function of the pdf of $x(t)$ asymptotically possessing the symmetry:
\begin{equation}
\label{eq-FRSymmetry}
I(x) - I(-x) = -\Sigma x.
\end{equation}
It is easy to show that Eq.~(\ref{eq-kramerFunction}), satisfies this symmetry exactly with a rate, $\Sigma$, given by 
\begin{equation}
\label{eq-Sigma}
\Sigma = \frac{2 \rho}{(1\!-\!\rho^2)\so\st}.
\end{equation}
From this, we conclude that when it is reasonable to model non-equilibrium fluctuations using a product of correlated  normal variables a FR will result. The value of the entropy rate, $\Sigma$, depends on the dissipation rate. This observation may partially explain the proliferation of empirical Fluctuation Relations in the literature and the lack of consensus on the value and meaning of the entropy rate measured for different experimental situations. This result is entirely statistical and does not require any restrictions on the microscopic dynamics such as time--reversibility. Indeed it tells us very little about the physics of the system under study.

\begin{figure}
\includegraphics[width=6.5cm]{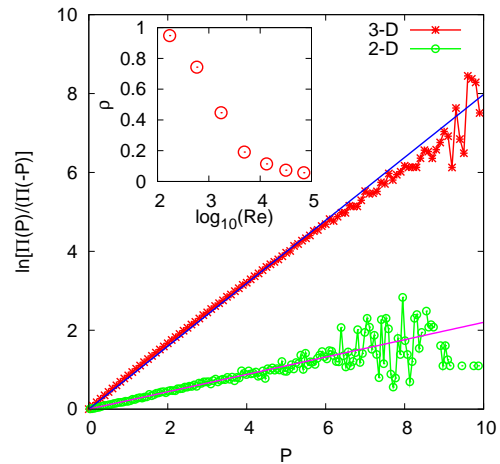}
\caption{\label{fig-FR}(Color online) Asymmetry of the pdfs of local power in 2-D and 3-D. Solid lines
indicate the prediction of Eq.~(\ref{eq-Sigma}). Inset shows the decrease of the correlation coefficient for the 2-D case, $\rho$, as the notional Reynolds number increases. We expect a similar trend in 3-D.}
\end{figure}

Let us now reconsider the specific case of turbulent power fluctuations. Fig.~\ref{fig-FR} shows the degree to which our numerical data satisfies the symmetry of Eq.~(\ref{eq-FRSymmetry}) with the appropriate values of $\Sigma$ from Eq.~(\ref{eq-Sigma}). As in many cases in the literature, a good agreement is found for relatively small fluctuations but a systematic deviation appears for very large fluctuations. Unusually, we understand completely the observed values of $\Sigma$. It is determined solely from the the correlation coefficient, $\rho$, which is not known a-priori. The inset of Fig.~\ref{fig-FR} shows numerical measurements of how $\rho$ varies as the notional Reynolds number, $Re$, is increased. In 2-D, in the presence of an inverse cascade, the usual definition of $Re$ is of questionable usefulness, since the principal energy balance is between nonlinearity and large scale dissipation. Nonetheless, it is widely used so we adopt it here to parameterise our simulations. We observe that $\rho$ decreases as $Re$ increases so that the pdf of the power becomes more symmetric as the flow becomes more turbulent. This make physical sense as the greater the turbulent fluctuations, the less the velocity can correlate with the forcing. As the pdf of the power becomes more symmetric, Eq.~\ref{eq-meanXY} demonstrates that the decrease in the correlation must be compensated for by an increase in the variance of the velocity field if one is to maintain a fixed mean rate of energy injection. There are clearly some important questions to address here in understanding the relationship between $\rho$ and $Re$ as well as investigating the corresponding issues in 3-D. These are, however, beyond the scope of the present work.

\begin{figure}
\includegraphics[width=6.5cm]{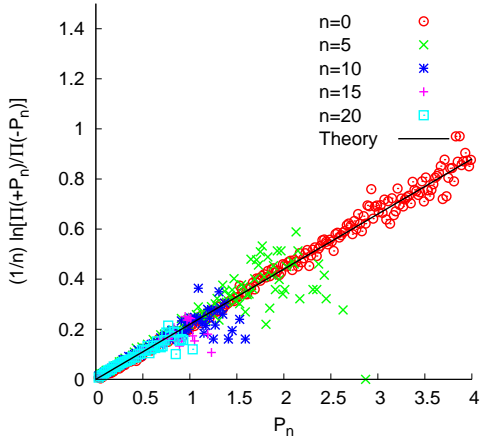}
\caption{\label{fig-lagrangianFR}(Color online) Fluctuation relation for the Lagrangian power averaged over time ($\tau$) intervals in multiples of the correlation time ($\tau_c$) of the power signal for $\tau/\tau_c \equiv n$ = 0, 5, 10, 15, and 20. The solid line is the theoretical prediction from Eq. \ref{eq-Sigma}. }
\end{figure}

We have already discussed how the Kramer function, Eq.~\ref{eq-kramerFunction}, encodes information about the behaviour of sums of samples from the pdf. If we think of local averaging as such a summation procedure, the fact that the  Kramer function exhibits a FR with a rate given by Eq.~\ref{eq-Sigma}, means that we might expect the coarse-grained power to satisfy this FR provided that we coarse-grain the data over intervals longer than the correlation length. This latter condition is important since the Kramer function describes the asymptotics of sums of {\em independent} samples. This provides a way to link our discussion of local power fluctuations to ``global'' fluctuations (in the sense of fluctuations at scales of many correlation lengths). This coarse-graining could be done either in space or in time. In our numerical simulations, the spatial correlation length was too long to allow us to perform the coarse-graining convincingly and will require further effort. This is unfortunate, this being most relevant to experiments. The Lagrangian correlation time is relatively much shorter. Therefore we can illustrate the point by coarse-graining temporally using data gathered from measurements of the force and velocity in the Lagrangian frame. Full details of the Lagrangian measurements are already available in \cite{BC2008}. We define a temporally coarse-grained power, $P_n(t)$, again normalised by the standard deviations of the force and velocity:
\begin{equation}
\label{nondimp}
P_n(t) = \frac{1}{\sigma_v \sigma_f} \frac{1}{n \tau} \int_t^{t+n \tau} P(t^\prime)\, dt^\prime.
\end{equation}
Here $\tau$ is the Lagrangian correlation time ($\tau\approx 0.1$ in our simulations compared with a large eddy turnover time of about 10) and $P(t)$ is the local power in the Lagrangian frame. The results for the pdfs of $P_n$ are shown in Fig.~\ref{fig-lagrangianFR} for coarse-graining times ranging from 5 to 20 correlation lengths. It is clear that the symmetry of the Kramer function demonstrated in Eq.~\ref{eq-FRSymmetry} produces a FR for the coarse-grained power.

It has been rightly argued \cite{AFMP2001,FLF2008} that the deviations from Eq.~(\ref{eq-FRSymmetry}) evident in Fig.~\ref{fig-FR} are typical. Here we understand that these deviations do not follow from the statistical model proposed in \cite{BC2008} and \cite{FLF2008} but rather are a signature of some underlying dynamics. For the specific case of turbulence, the work of \cite{FL1997} identified specific flow configurations (``instantons'') which are responsible for the faster--than--Gaussian decay of the single point velocity distribution  in forced turbulence. This theory may provide a starting point for analysis of the deviations from the FR observed in our data but given the non-universal nature of the force, it seems unlikely that there is anything universal about these deviations.

\noindent{\bf Acknowledgements}

This work was partially carried out under the auspices of the National Nuclear Security
Administration of the U.S. Department of Energy at Los Alamos National
Laboratory under Contract No.  DE-AC52-06NA25396.

\bibliography{main}

\end{document}